\pdfoutput=1

\documentclass[prd,amsmath,amssymb,floatfix,nofootinbib,superscriptaddress, twocolumn]{revtex4}

\usepackage{bm}
\usepackage{amsmath}
\usepackage{epsfig}
\usepackage{color}
\usepackage{natbib}
\usepackage{graphicx}
\usepackage{enumerate}
\usepackage{natbib}
\usepackage[hypertex]{hyperref}
\usepackage{ifthen}

\def\eprinttmp@#1arXiv:#2 [#3]#4@{
\ifthenelse{\equal{#3}{x}}{\href{http://arxiv.org/abs/#1}{#1}}{\href{http://arxiv.org/abs/#2}{arXiv:#2} [#3]}}

\providecommand{\eprint}[1]{\eprinttmp@#1arXiv: [x]@}
\newcommand{\adsurl}[1]{\href{#1}{ADS}}

\def\be{\begin{equation}}
\def\ee{\end{equation}}
\def\ba{\begin{eqnarray}}
\def\ea{\end{eqnarray}}

\def\nn{\nonumber}

\renewcommand{\ell}{l}
\newcommand{\slm}[4]{{}_{#2} {#1}_{#3 #4}}

\newcommand{\yslm}[3]{\slm{Y}{#1}{#2}{#3}}
\newcommand{\degree}{^{\rm o}}
\newcommand{\clo}{{\cal O}}

\newcommand{\threej}[6]{\left(
                           \begin{array}{ccc}
        \! #1\! & #2\!  & #3\!  \\
        \! #4\! & #5\!  & #6\!
                           \end{array}
                   \right)}

\begin{document}

\title{Asymmetric Beams and CMB Statistical Anisotropy}

\author{Duncan Hanson}
\affiliation{Institute of Astronomy and Kavli Institute for Cosmology Cambridge, University of Cambridge, Madingley Road, Cambridge CB3 OHA, United Kingdom}

\author{Antony Lewis}
\affiliation{Institute of Astronomy  and Kavli Institute for Cosmology Cambridge, University of Cambridge, Madingley Road, Cambridge CB3 OHA, United Kingdom}
\affiliation{Department of Physics \& Astronomy, Pevensey II Building, University of Sussex, Falmer, Brighton BN1 9QH, UK}

\author{Anthony Challinor}
\affiliation{Institute of Astronomy  and Kavli Institute for Cosmology Cambridge, University of Cambridge, Madingley Road, Cambridge CB3 OHA, United Kingdom}
\affiliation{DAMTP, Centre for Mathematical Sciences, University of Cambridge, \\ Wilberforce Road, Cambridge CB3 OWA, UK}

\begin{abstract}
\baselineskip 11pt
Beam asymmetries result in statistically anisotropic cosmic microwave background (CMB) maps.
Typically, they are studied for their effects on the CMB power
spectrum, however they more closely mimic anisotropic effects
such as gravitational lensing and primordial power asymmetry.
We discuss tools for studying the effects of beam asymmetry on
general quadratic estimators of anisotropy,
analytically for full-sky observations as well as in the analysis of
realistic data.
We demonstrate this methodology in application to a recently detected $9\sigma$
quadrupolar modulation effect in the WMAP data, showing that beams provide
a complete and sufficient explanation for the anomaly. 
\end{abstract}
\maketitle

The cosmic microwave background (CMB) is a powerful
probe for both modern and future cosmology: its rotationally invariant
power spectra have been instrumental in hammering
out the details of the current concordance model, its
non-Gaussianities have the potential to discriminate
between various early-universe models, 
and its statistical anisotropies can be used to probe astrophysically interesting
secondary effects such as gravitational lensing. 

We observe the CMB through the convolution of an instrumental
beam, an effect which must be carefully treated in analysis.
Qualitatively, the effects of beams are twofold:
(i) they suppress structures on scales smaller than the beam size; and
(ii) if the beams are asymmetric, they can introduce statistical anisotropies
into the observed CMB which can bias estimators for other anisotropic signals.
%
The purpose of this paper is to collect results on the simulation of
beam effects, and to present fast, accurate techniques for
forecasting and correcting the effects of beams on estimators of statistical anisotropy.
In Sec.~\ref{sec:model} we present a model of beam asymmetries and
 we discuss the covariance which beams produce
in the observed CMB in Sec.~\ref{sec:covariance}. 
In Sec.~\ref{sec:qe} we derive the effects of the anisotropic covariance
on anisotropy
estimators, and in Sec.~\ref{sec:case_study} we illustrate this discussion by applying
these techniques to study the effects of beams on the highly significant
quadrupolar modulation effect in the WMAP data. Our
conclusions are collected in Sec.~\ref{sec:conclusions}.
The effect of beam asymmetries on the estimated power spectrum of the
CMB for general survey geometries is discussed in
Appendix~\ref{sec:beam_power}.

\section{Model}
\label{sec:model}

In a realistic CMB observation, the effective sky signal at each point
in the time-ordered data (TOD) is a convolution of the true sky signal
with the experimental beam, oriented according to the scan strategy.
Schematically, we have
\be
T_i = \int_{S^2} d \Omega\, r_{i}(\Omega) \Theta(\Omega) + n_i
\label{eqn:obs_real}
\ee
where $T_i$ is the temperature for time-step $i$ in the TOD,
$\Theta(\Omega)$ is the underlying CMB signal,
$r_{i}(\Omega)$ is the beam response and $n_i$ is the instrumental noise.
For the purposes of compact notation we will abbreviate $\int_{S^2} d \Omega$ as $\int$
for the remainder of this paper.
The integral in Eq.~\ref{eqn:obs_real}
can be performed 
by brute force in real-space
using interpolation on pixelized maps of the beam and sky  \cite{Wehus:2009zh}.
For this approach to be computationally feasible, the beam
must be assumed zero outside some small patch where its
response is peaked, and so it is difficult to study sidelobe
effects with this approach, although it can be quite fast.

In this work, we will find it more useful to work in harmonic
space, where the effects of beams are easier to study analytically.
We begin by writing the beam response as a harmonic sum.
If we center the beam at the north pole, with some fiducial beam axis aligned along
the $+x$-axis (the $\phi=0$ meridian), and expand it in spherical harmonics $b_{\ell m}$,
then the beam at location $\Omega$ for the $i^{\rm th}$ TOD observation is given by
(e.g. \cite{Souradeep:2001ds})
\be
r_{i}(\Omega) =
\sum_{s=-s_{\rm max}}^{s_{\rm max}}
\sum_{\ell=|s|}^{\ell_{\rm max}}
\sum_{m=-\ell}^{\ell}
D^{\ell}_{m s}(\phi_i, \theta_i, \alpha_i) b_{\ell s} \, \yslm{0}{\ell}{m}(\Omega).
\label{eqn:beam}
\ee
For the purposes of compact notation we will
drop the summation limits in what follows.
The limits themselves will be discussed later.
The action of the Wigner-$D$ matrix can be visualized as follows:
imagine fixing the coordinate system in space and performing
right-handed rotations of the beam image
about the $z$-axis by an angle $\alpha_i$,
then about the $y$-axis by an angle $\theta_i$, and finally
about the $z$-axis by an angle $\phi_i$.
The first rotation gives the beam its orientation: $\alpha_i$ is the angle of the
fiducial beam axis, measured from the southern side of 
the meridian which passes through the pixel location $(\theta_i,
\phi_i)$ assigned to the observation.
We use $\yslm{s}{\ell}{m}$ to denote a spin-weighted spherical harmonic,
of which the standard spherical harmonics are a special case with
$s=0$. Unless otherwise noted, the harmonics should be taken as
functions of $\Omega$.
We can then rewrite Eq.~\eqref{eqn:obs_real} as
\ba
T_i
&=& \sum_{\ell m s} D^{\ell}_{-m s}(\phi_i, \theta_i, \alpha_i) b_{\ell s} (-1)^{m} \Theta_{\ell m} + n_i \nn \\
&=& \sum_{\ell m s} e^{-i s \alpha_i} B_{\ell s} \Theta_{\ell m} \, \yslm{s}{\ell}{m}(\theta_i, \phi_i) + n_i.
\label{eqn:obs}
\ea
In the second step we have used the close relationship between Wigner-$D$ functions
and the spin-weighted spherical harmonics \cite{VarQuant}, and
introduced the beam transfer function
$B_{l s}$ given by
\be
B_{\ell s} = \sqrt{ \frac{4\pi}{2l+1} } b_{ls}.
\ee
For a beam which is normalized to have unit response to a monopole,
$B_{00} = 1$. We shall refer to the $s\!=\!0$ coefficients of the beam as the symmetric part,
as they represent the component of the beam which depends only on the radial
distance from its center.
The $s\!\ne\!0$ coefficients encapsulate beam asymmetry.
On scales much smaller than the beam size, $B_{ls}$ becomes very
small, which effectively band-limits $B_{ls} \Theta_{lm}$.
This scale determines $\ell_{\rm max}$.
The evaluation of Eq.~\eqref{eqn:obs} may then be performed in
$\clo(s_{\rm max} l_{\rm max}^2)$.
To perform this convolution for each timestep in the TOD is in general
prohibitively expensive, and some approximations must be made.
There are several possible approaches:

$\mathbf{(1)}$
The convolution can be computed over a grid covering the full rotation group
with fast-Fourier-transforms for the Euler angles $\phi$ and $\alpha$ and, optionally, for
$\theta$~\cite{Wandelt:2001gp}. The TOD can be obtained by interpolation off
this grid.
%

$\mathbf{(2)}$ %
If the beam may be represented using a
small number of symmetric basis functions, each of
these may be rapidly convolved in harmonic space and
then sampled based on the location and orientation of
these basis functions for each sample in the TOD \cite{Tristram:2003ee}.
The difficulty here is the ability to represent the beam as a sum of
symmetric functions. Note that in the limit that
the beam is represented as a sum of delta
functions, this approach is conceptually the same as
real-space integration.

We note the above approaches for completeness.
In this work, we will use a popular \cite{Smith:2007rg,Hirata:2004rp,Hirata:2008cb,Mitra:2004nx} map-based approach based on the assumption that the TOD noise is
uncorrelated on the timescales which separate pixel
visits. In this case, it is a good approximation to
the mapmaking process (in the absence of beam deconvolution \cite{Armitage:2004pk}) to take
\be
\tilde{\Theta}(\Omega_p) + n(\Omega_p) = \sum_{ i \in p} T_i / H_p,
\label{eqn:beam_asym_map}
\ee
where $\tilde{\Theta}$ is an effectively observed sky, $n(\Omega_p)$ is a noise map, 
and $H_p$ is the number of elements in the sum, which is taken over all hits assigned to pixel $p$, with center at $\Omega_p$.
This approach can also be used for differencing experiments,
which suppress correlations between TOD samples by mapmaking
from the difference between two nearly-identical detectors, to
remove common mode fluctuations. In this case, one can use
an effective beam which is a hit-weighted sum of the two beams
which are differenced \cite{Hinshaw:2006ia}.

In conjunction with Eq.~\eqref{eqn:obs}, the sum of Eq. \eqref{eqn:beam_asym_map} can be seen to effect a Fourier
transform of the distribution of orientation angles, with an effective observed
sky given by \cite{Smith:2007rg}
\be
\tilde{\Theta}(\Omega_p)  = \sum_{s} w(\Omega_p,-s) \left[ \sum_{\ell m} B_{\ell s}  \Theta_{\ell m}
\, \yslm{s}{\ell}{m}(\Omega_p) \right],
\label{eq:obs_sky}
\ee
where the details of the scan strategy are contained in the spin $-s$ field
\be
w(\Omega_p,-s) = \sum_{i \in p} e^{-i s \alpha_i} / H_p.
\label{eq:spin_weight}
\ee
Since ${}_s Y_{lm}$ involves $s$ (spin-weighted) derivatives of the $Y_{lm}$,
each term in the $s$ sum is the real-space product of the
scan strategy and beam-filtered derivatives of the CMB.
For a beam which is approximately azimuthally symmetric, or a scan strategy
which broadly distributes the orientation angles, $w(\Omega_p,-s) B_{\ell s}$ falls off sharply
with $s$, and it follows that calculation of only the lowest $s$ terms
suffice to give a good approximation to the beam-convolved map.
This determines an effective $s_{\rm max}$ which can be much less
than that naively required to describe accurately the beam in Eq.~\eqref{eqn:beam}.

Given a scan strategy, Eq.~(\ref{eq:obs_sky}) provides an $\clo(s_{\rm max} l_{\rm max}^3)$ method
to compute effectively the sky observed by an experiment
with an asymmetric beam $B_{ls}$ and given scan strategy
$w(\Omega_p,s)$. This approximation is useful not only for its speed,
but also to gain an intuitive analytical understanding of beam effects, which we proceed
to discuss in the following sections.

\section{Covariance}
\label{sec:covariance}
Beam effects are linear in the underlying CMB,
and so do not affect its (assumed) Gaussianity.
For Gaussian models, the statistics of the observed CMB remain completely
characterized by its covariance. The effect of
beams is simply to introduce statistical anisotropies
which give off-diagonal and $m$-dependent contributions
to the covariance.

In harmonic space the beam-convolved sky is given by
\begin{align}
\tilde{\Theta}_{\ell' m'}
&= \sum_{L M S} \sum_{\ell m}  B_{\ell S}\, \Theta_{\ell m} \,\slm{w}{-S}{L}{M}
\nonumber \\
&\mbox{} \times
\int \yslm{-S}{L}{M} \yslm{0}{\ell'}{m'}^{*} \yslm{S}{\ell}{m} ,
\label{eqn:obs_eff_har}
\end{align}
where
\be
\slm{w}{S}{L}{M} = \int w(\Omega, S) \yslm{S}{L}{M}^*
\ee
are the spin-$S$ multipoles of $w(\Omega,S)$.

The covariance of the beam-convolved CMB may then be written as
\ba
\tilde{C}_{\ell_1 m_1\, \ell_2 m_2}
&=& \langle \tilde{\Theta}_{\ell_1 m_1} \tilde{\Theta}_{\ell_2 m_2}^{*} \rangle \nn \\
&=& \delta_{\ell_1 \ell_2} \delta_{m_1 m_2} B_{\ell_1 0}^{2}
C^{\Theta\Theta}_{\ell_1} \nn \\
&& \quad + (-1)^{m_2} \Delta_{\ell_1 m_1\,\ell_2 -m_2}.
\ea
The $\Delta_{\ell_1 m_1\, \ell_2 m_2}$ term contains the part of the covariance which is due to
beam asymmetries. We further split it into two terms, such that
$\Delta = \Delta^{(1)} + \Delta^{(2)}$.
The $\Delta^{(1)}$ terms are those which couple
an $s\!=\!0$ mode of the convolution with an $s\!\ne\!0$ mode.
The notation arises because we think of them as being first order
in any beam asymmetry:
\begin{multline}
\Delta^{(1)}_{\ell_1 m_1\, \ell_2 m_2}
=
\sum_{S \ne 0} \sum_{LM} \slm{w}{S}{L}{M}
 \\\times \int \yslm{S}{L}{M}^*
    \biggl[ \yslm{0}{\ell_1}{m_1}\, \yslm{S}{\ell_2}{m_2} \,
  B_{l_2 S}^* B_{l_2 0} C^{\Theta \Theta}_{\ell_2} 
  + (1 \leftrightarrow 2) \biggr].
\label{eqn:delta_1}
\end{multline}
The $(1 \leftrightarrow 2)$ represents the interchange of $\ell_1, \ell_2$ and $m_1, m_2$ 
in the preceding expression. The $\Delta^{(2)}$ terms couple two $s\!\ne\!0$ modes:
\begin{multline}
\Delta^{(2)}_{\ell_1 m_1\, \ell_2 m_2}
=
\sum_{ \substack{S_1 \ne 0 \\ S_2 \ne 0} }
\sum_{lm} (-1)^{m}
B_{\ell S_1} B_{\ell S_2} C^{\Theta \Theta}_{\ell}
\\ 
\times \sum_{L_1 M_1} {}_{-S_1} w_{L_1 M_1}
\left( \int \yslm{-S_1}{L_1}{M_1}
  (\yslm{0}{\ell_1}{m_1}^{*}) \yslm{S_1}{\ell}{-m} \right)
\\
\times \sum_{L_2 M_2} {}_{-S_2} w_{L_2 M_2}
\left( \int \yslm{-S_2}{L_2}{M_2}
  (\yslm{0}{\ell_2}{m_2}^{*}) \yslm{S_2}{\ell}{m} \right) .
\label{eqn:delta_2}
\end{multline}

Traditionally, CMB analyses have focused on the power spectrum,
or average diagonal elements of the covariance matrix for each $l$.
The $\Delta^{(1)}$ term evaluates to zero for these elements,
however, and so the effects of beam asymmetries on
the power spectrum are due solely to the $\Delta^{(2)}$ terms.
For estimators of statistical anisotropy,
however, the dominant contributions to
$\Delta$ are expected to be the $\Delta^{(1)}$ terms, which involve
only one power of $B_{\ell (s\ne0)}$.
Beam asymmetries therefore have quantitatively different
effects on the power spectrum and on anisotropy estimators.
A further important difference is that beam asymmetries are only
an issue for the power spectrum on small scales, but, due to mode coupling,
they can still be important when reconstructing large-scale modes
of any statistical anisotropy.
Note that the above arguments only apply exactly on the full-sky
with uniform pixel weighting. For
pseudo-$C_l$ power spectra on a cut sky or with anisotropic weighting
(for example, to mitigate inhomogeneous pixel noise),
the beam anisotropies can couple with the asymmetry introduced
by the weights, which gives $\Delta^{(1)}$ a contribution
to the power spectrum. However, this is only significant near strong
inhomogeneities in the pixel weights, and is therefore generally suppressed.
For the remainder of this paper, we will focus on the effects of
beams on anisotropy estimators. The effects
on the power spectrum are discussed in Appendix~\ref{sec:beam_power};
see also~\cite{Hinshaw:2006ia}.

\section{Anisotropy Estimators}
\label{sec:qe}
The CMB is assumed to be statistically isotropic to a good approximation, but there may be small contributions to the covariance from a variety of effects, such as gravitational lensing (if the lensing potential is considered as fixed, see e.g. \cite{Lewis:2006fu} for a review), inhomogeneous reionization (if the reionization history is fixed, see e.g. \cite{Dvorkin:2008tf}), Doppler modulation \cite{Challinor:2002zh}, or more exotic statistical anisotropy (e.g. \cite{Hanson:2009gu} and references therein). Following the notation introduced in the previous section, we will write the CMB covariance as
\be
\langle \Theta_{\ell_1 m_1} \Theta_{\ell_2 m_2}^* \rangle = \delta_{l_1 l_2}\delta_{m_1 m_2}C_{l_1}^{\Theta \Theta} + \sum_i (-1)^{m_2} \Delta^{(i)}_{l_1 m_1\, l_2 -m_2},
\label{eq:aniso_cov}
\ee
where $i$ labels the various physical effects that contribute to the
anisotropy. 

If we assume that the anisotropy from each effect $i$ is sourced linearly
by multipoles ${}_{S^i} f^{(i)}_{LM}$ with spin-weights $\{ S^i \}$ that
satisfy $[{}_{S^i} f^{(i)}_{LM}]^* = (-1)^{S^i+M} {}_{-S^i} f^{(i)}_{L\, -M}$,
then covariance under rotations 
(i.e. the requirement that if $\Theta$ is rotated, $f_{LM}$ must rotate in tandem)
and parity (${}_{S^i}f^{(i)}_{LM}
\rightarrow (-1)^{S^i+L} {}_{-S^i}f^{(i)}_{LM}$)
generally requires that 
each term in Eq.~(\ref{eq:aniso_cov}) has the form
\begin{multline}
\Delta_{l_1 m_1\, l_2 m_2}^{(i)}  = 
\sum_{S^i LM} \slm{f}{S^i}{L}{M}^{(i)}  \\
\times \sum_{s_1}\left( \int \yslm{S^i}{L}{M}^*
\yslm{s_1}{\ell_1}{m_1}
\yslm{s_2}{\ell_2}{m_2}
\right)
W_{S^i,s_1}^{(i)}(\ell_1, \ell_2, L) ,
\label{eqn:cov}
\end{multline}
where $W_{S^i,s_1}^{(i)}(\ell_1, \ell_2, L)=(-1)^{S^i}W_{-S^i,-s_1}^{(i)}(\ell_1,
\ell_2,L)$ is a weight function 
which describes the way in which the anisotropy field couples the $\Theta$ multipoles, while
%
$s_1$ and $s_2 \equiv S^i-s_1$ label different partitions of the spin
$S^i$ between two spin-weighted harmonics
\footnote{Two examples may help to solidify the notation.
For lensing, $\Theta \rightarrow \Theta + d^a \nabla_a \Theta$ to
first order in the lensing deflection $d^a$. This gives a covariance
with nonzero weights
\begin{align*}
W^{(\text{lens})}_{\pm1,\pm 1} (l_1,l_2,L) &= \mp \frac{C_{l_1}^{\Theta\Theta}}{2\sqrt{l_1(l_1+1)}}\\
W^{(\text{lens})}_{\pm1,0} (l_1,l_2,L) &= \mp \frac{C_{l_2}^{\Theta\Theta}}{2\sqrt{l_2(l_2+1)}} ,
\end{align*}
and $\{{}_{S}f^{(\text{lens})}_{LM}\} = \{{}_{\pm 1} d_{LM}\}$ are the spin-weight
multipoles of the deflection field. Note that we have not assumed that
$d^a$ is a gradient here. For beam asymmetries with
${}_{S}f^{(\text{beams})}_{LM} = {}_S w_{LM}$, the nonzero weights are
$W^{(\text{beams})}_{S,0}(l_1,l_2,L)=B_{l_2 S}^* B_{l_2 0} C_{l_2}^{\Theta\Theta}$ and
$W^{(\text{beams})}_{S,S}(l_1,l_2,L)=B_{l_1 S}^* B_{l_1 0} C_{l_1}^{\Theta\Theta}$.
}.
Typically, one of $s_1$ or $s_2$
is zero since we are dealing here with the
spin-0 temperature.
It can be seen that the $\Delta^{(1)}$ term of the beam covariance
in the previous section is of this form (although the smaller $\Delta^{(2)}$ term is not).
The $\Delta^{(i)}_{\ell_1 m_1\, \ell_2 m_2}$ are symmetric under $(1\leftrightarrow
2)$ so we may take $W^{(i)}_{S^i,s_1}(l_1,l_2,L) = W^{(i)}_{S^i,s_2}(l_2,l_1,L)$.
Moreover, $(-1)^{m_2} \Delta^{(i)}_{\ell_1 m_1\, \ell_2 -m_2}$ is Hermitian
under $(1\leftrightarrow 2)$ which gives rise to the spin-flip symmetry
\be
W^{(i)}_{S^i,s_1}(l_1,l_2,L) = (-1)^{S^i}W^{(i)\ast}_{-S^i,-s_2}(l_2,l_1,L) .
\ee
Note that the quantity $W(\ell_1, \ell_2, L) \slm{f}{S}{L}{M}$
is essentially equivalent to the bipolar spherical
harmonic coefficients, $A_{l_1 l_2}^{LM}$, of
Hajian and Souradeep~\cite{Hajian:2003qq}. 
The formalism which we will use here and the bipolar spherical
harmonic formalism can be thought of as two different representations
of the same symmetry relations, analogous to e.g. Clebsch-Gordan coefficients and
Wigner $3j$ symbols. Similarly, their relative benefits depend on
the use. For calculations, the quadratic estimator approach
often results in simpler expressions, but the $A_{\ell_1 \ell_2}^{LM}$ can
prove more useful for blind searches, or for gaining insight into the
physical interpretation of the anisotropies and
the relationships between different models \cite{Bennett:2010jb}.

As discussed in \cite{Hanson:2009gu}, optimal quadratic maximum-likelihood estimators
can be constructed for $\slm{f}{S}{L}{M}$, under the assumption that their effects are perturbative.
The estimators approximately maximize the Gaussian log-likelihood $\cal{L}$ with respect to the $\slm{f}{S}{L}{M}$, 
so that they solve $\partial{\cal L}/\partial \slm{f}{S}{L}{M} = 0$. 
The solution is constructed from a set of
quadratic building blocks $\slm{h}{S^i}{L}{M}^{(i)}$, each of the form
%
\ba
\slm{h}{S^i}{L}{M}^{(i)} &=& \frac{1}{2} \sum_{\ell_1 m_1, \ell_2 m_2}
\bar{\Theta}_{\ell_1 m_1}
\left(\frac{\partial \Delta^{(i)}_{\ell_1 m_1\, \ell_2 m_2}}
{\partial {}_{S^i}f^{(i)}_{LM}}\right)^*
\bar{\Theta}_{\ell_2 m_2}
\nonumber \\
&=& \frac{1}{2}
\sum_{\ell_1 m_1, \ell_2 m_2, s_1}
\left[ \int \yslm{S^i}{L}{M}^* \yslm{s_1}{\ell_1}{m_1} \yslm{s_2}{\ell_2}{m_2} \right]
\nn \\ && \qquad \quad \quad \times
W^{(i)\ast}_{S^i,s_1}(\ell_1, \ell_2, L) \bar{\Theta}_{\ell_1 m_1} \bar{\Theta}_{\ell_2 m_2},
\label{eqn:est}
\ea
where $\bar{\Theta}_{\ell m}$ is the inverse-variance filtered observed sky (in general the observed sky premultiplied by the signal-plus-noise inverse covariance including all anisotropic contributions to the covariance that do not depend on the set of parameters that are being estimated).
The inverse-variance filtering can be performed quickly using conjugate descent
with a good preconditioner, the best to date being that of \cite{Smith:2007rg}.
The estimator for $\slm{f}{S^i}{L}{M}^{(i)}$ is then given by
\ba
\slm{\hat{f}}{S^i}{L}{M}^{(i)} = \sum_{L' M' j S^j} {\cal F}^{-1}_{iS^iLM, jS^jL'M'}
\left[ \slm{h}{S^j}{L'}{M'}^{(j)} - \langle \slm{h}{S^j}{L'}{M'}^{(j)} \rangle \right],
\ea
where the ensemble average is taken over realizations of the CMB and noise. The ``mean-field'' term $\langle \slm{h}{S^i}{L}{M}^{(i)} \rangle$ subtracts off anisotropy due to anisotropic noise, sky cuts, and known anisotropic components of the covariance (e.g. beam asymmetries). The matrix
${\cal F}^{-1}$ is the inverse of the Fisher matrix, which is calculated as
\begin{multline}
{\cal F}_{i S^i LM, j S^j L'M'} 
=
\frac{1}{2} \sum_{\ell_1 m_1, \ldots , \ell_4 m_4}
(-1)^{m_1+m_2} C^{-1}_{\ell_1 m_1\,\ell_2 m_2} 
\\
\hspace{-2cm} \times \left(
\frac{\partial \Delta^{(i)}_{\ell_3 m_3\, \ell_2 -m_2}}{\partial {}_{S^i}
f^{(i)}_{LM}}\right)^* C^{-1}_{\ell_3 m_3\, \ell_4 m_4}
\frac{\partial \Delta^{(j)}_{\ell_4 m_4\, \ell_1 -m_1}}{\partial {}_{S^j}
f^{(j)}_{L'M'}} ,
\end{multline}
where $C^{-1}$ is the inverse covariance matrix used to construct
$\bar{\Theta}_{lm}$. The Fisher matrix
can be shown to equal the covariance of the $\slm{h}{S^i}{L}{M}^{(i)}$:
\be
{\cal F}_{i S^i LM, j S^j L'M'}
= \langle \slm{h}{S^i}{L}{M}^{(i)} \slm{h}{S^j}{L'}{M'}^{(j)*} \rangle
-
\langle \slm{h}{S^i}{L}{M}^{(i)} \rangle  \langle \slm{h}{S^j}{L'}{M'}^{(j)*} \rangle .
\ee
For an observation and underlying CMB that are statistically
isotropic (e.g. full-sky coverage with homogeneous noise levels
and symmetric beams), rotational symmetry requires that the inverse-variance
filtered CMB has a diagonal covariance
\be
\langle \bar{\Theta}_{\ell_1 m_1} \bar{\Theta}^*_{\ell_2 m_2} \rangle
=
 \frac{\delta_{\ell_1 \ell_2}\delta_{m_1 m_2}}{C^{\rm{tot}}_{\ell_1}} \quad \rm{(iso.)},
\ee
where $1/C^{\rm{tot}}_{\ell}$ is the inverse-variance filter.
This propagates to the Fisher matrix ${\cal F}$, which is then also diagonal in $L$,
and independent of $M$:
\begin{multline}
{\cal F}_{iS^i LM, j S^j L'M'}^{\rm iso.} = \delta_{LL'} \delta_{M M'}
\\
\times \sum_{\ell_1 \ell_2 s_1 s_1'}(-1)^{S^i+S^j}
\frac{(2\ell_1+1)(2\ell_2+1)}{8\pi C^{\rm{tot}}_{l_1} C^{\rm{tot}}_{l_2}}
W_{S^i,s_1}^{(i)\ast}(\ell_1, \ell_2, L) 
\quad \quad \\ 
\times \ W_{S^j,s_1'}^{(j)}(\ell_1, \ell_2, L) 
\threej{\ell_1}{\ell_2}{L}{-s_1}{-s_2}{S^i}
\threej{\ell_1}{\ell_2}{L}{-s_1'}{-s_2'}{S^j}. \quad \quad
\end{multline}
The isotropic Fisher matrix is chiefly useful for forecasting purposes.
In practice, inhomogeneous sky coverage and foreground cuts
mean that it should be estimated from simulations.

If the weights are separable, such that $W(\ell_1, \ell_2, L) = W_1(\ell_1) W_2(\ell_2) W_L(L)$,
or can be decomposed as a sum of separable terms, then these estimators have fast
position-space forms with computational cost $\clo(\ell_{\rm max}^3)$, and the isotropic
Fisher matrix can be evaluated in ${\cal O}(\ell_{\rm max}^2)$ \cite{Dvorkin:2008tf}.

It can be seen clearly from Eq.~\eqref{eqn:delta_1} that
these tools apply to the covariance produced by beam asymmetries.
Optimal estimators could, for example, be formed to reconstruct
the components of $\slm{w}{S}{L}{M}$ for each $S$.
In practice, the instrumental scan strategy is fixed and asymmetric beams act
as a source of bias for other anisotropy estimators, which have
the form of Eq.~\eqref{eqn:est}. In this view, beams simply make a
contribution to the covariance of the observed sky. They should be incorporated into
the inverse-variance filtering operation and the mean-field
subtraction. For realistic observations, this can be done
easily as the inverse-variance filtering step is done using
conjugate descent, and requires only a fast method to apply
the beam effects, such as that provided by Eq.~\eqref{eqn:obs_eff_har}. This is demonstrated
at the TOD level in \cite{Armitage:2004pk}, for example.
The mean-field can be determined straightforwardly
from simulations. 
Analytic calculations are also feasible if the inverse-variance filter is isotropic,
and can be useful for forecasting purposes.
Neglecting a known source of anisotropy in the data during mean-field
subtraction will generally bias anisotropy estimators for other effects.
Explicitly, the mean-field bias on an estimator $\slm{\hat{f}}{S^j}{L}{M}^{(j)}$
with weight function $W_{S^j,s_1}^{(j)}(\ell_1, \ell_2, L)$ by a contaminant $i$
with covariance as in Eq.~\eqref{eqn:cov} is
\begin{multline}
\langle \slm{h}{S^j}{L}{M}^{(j)} \rangle =
\slm{f}{S^i}{L}{M}^{(i)} \sum_{\ell_1 \ell_2 S^i s_1 s_1'} (-1)^{S^i + S^j}
\frac{(2\ell_1+1)(2\ell_2+1)}{8\pi C^{\rm{tot}}_{\ell_1} C^{\rm{tot}}_{\ell_2}}
\\
\times 
W_{S^i,s_1}^{(i)}(\ell_1, \ell_2, L)
\threej{\ell_1}{\ell_2}{L}{-s_1^{\ }}{-s_2^{\ }}{S^i}
\\
\times W^{(j)\ast}_{S^j,s_1'}(\ell_1, \ell_2, L)
\threej{\ell_1}{\ell_2}{L}{-s_1'}{-s_2'}{S^j} .
\label{eqn:qe_bias_general}
\end{multline}
A nice feature of this result is that the bias rather
directly traces the spatial distribution of the contaminant:
the bias is an isotropically filtered version of the contaminant.
In the case of pixel-uncorrelated anisotropic
instrumental noise, for example, the mean-field simply
traces the anisotropic part of the noise variance
map \cite{Hanson:2009dr}.
In the case of beams, on the other hand, the mean-field traces the components
of the scan strategy
$\slm{w}{S}{L}{M}$.

\section{Case Study: Primordial Power Asymmetry}
\label{sec:case_study}

As an example of this machinery in action,
we study a high-significance anomaly in the WMAP
data, which resembles the effects of a modulation
of the primordial power spectrum. Explicitly, one can
construct estimators based on the covariance for a $k$-space modulation
of the primordial power spectrum $ {\cal P}_{\chi}(\mathbf{k})$ 
with the form
\be
{\cal P}_{\chi}(\mathbf{k}) = {\cal P}_{\chi}(k)[1+g(\hat{\mathbf{k}})] .
\ee
If we take the bi-Copernican hypothesis that the universe has no preferred
orientation then the expectation is that $g(\hat{\mathbf{k}}) = 0$.
However current analyses of the WMAP maps strongly favor
a model in which $g(\hat{\mathbf{k}})$ has quadrupolar
components
\be
g(\hat{\mathbf{k}}) = \sum_{|M|\leq 2} g_{2M} \, \yslm{0}{2}{M}(\hat{\mathbf{k}}).
\ee
Furthermore, the preferred $g_{2M}$ are planar,
with $g_{2M} \propto \delta_{M0}$ in ecliptic coordinates  \cite{Hanson:2009gu,Groeneboom:2009cb,Bennett:2010jb}:
\be
g(\hat{\mathbf{k}}) = g_{20} \frac{1}{2} \sqrt{\frac{5}{4\pi}} (3 \cos^2
\theta_{\hat{\mathbf{k}}} - 1),
\ee
where $\theta_{\hat{\mathbf{k}}}$ is the angle from the ecliptic pole.
This form resembles the model proposed by \cite{Ackerman:2007nb}
(and other authors, e.g. \cite{Boehmer:2007ut})
which was the motivation for original detection made in \cite{Groeneboom:2008fz}.
A missing factor in the original version of  \cite{Ackerman:2007nb} resulted in the  ecliptic orientation being obscured; however,
this was corrected by \cite{Hanson:2009gu}.
The ecliptic alignment of the detected effect strongly suggests
an instrumental systematic or solar-system origin.
The signal is present at $9\sigma$ in the $W$ band \cite{Groeneboom:2009cb},
but varies strongly between detectors at the same
frequency \cite{Hanson:2009gu,Bennett:2010jb}, which singles out
an instrumental explanation, although
\cite{Groeneboom:2009cb} have also checked
the contribution of zodiacal light and found a
negligible effect.

Here we will continue the work of \cite{Hanson:2009gu}, using
optimal quadratic maximum-likelihood estimators to study the primordial
modulation effect. These estimators are often favorable to the
Gibbs-sampling approach of \cite{Groeneboom:2008fz,Groeneboom:2009cb}
for their speed, and the ease with which they can be modified to test
various systematic effects \cite{Hanson:2009gu,Bennett:2010jb}.
In the current application, the
quadratic estimator compresses millions of correlations
between thousands of observed modes to a small
handful of parameters, and should be effectively indistinguishable
from an exact likelihood analysis \cite{Bennett:2010jb}.
In the formalism of the previous section, the quadratic estimator
for $g_{2M}$ has $S=s_1=0$, and the weight function can be written as
\cite{Pullen:2007tu,Hanson:2009gu}
\be
W(\ell_1, \ell_2, 2) = \frac{ i^{\ell_1 - \ell_2} + i^{\ell_2 -
    \ell_1} }{2} C_{\ell_1 \ell_2},
\ee
where the $C_{\ell_1 \ell_2}$ matrix is given by
\be
C_{\ell_1 \ell_2} = 4\pi \int d \ln k {\cal P}_{\chi}(k)
\Delta_{\ell_1}(k) \Delta_{\ell_2}(k).
\ee
The $\Delta_{\ell}(k)$ used here are the angular CMB
transfer functions.

The largest expected instrumental effects which can produce
ecliptic-aligned anomalies are
inhomogeneous pixel noise levels and beam asymmetries.
The pixel-uncorrelated component of the instrumental
noise is already accounted for in current analyses, and
it has been argued that these estimators are insensitive
to percent-level changes in the noise amplitude \cite{Groeneboom:2009cb}.
We agree with this result: the mean-field for the quadratic
estimator of primordial power modulation due to WMAP noise
inhomogeneities is less than $1\sigma$ for all $V$- and $W$-band differencing assemblies (DAs), and so
percent-level changes in the noise level do not have appreciable effects.
In the bipolar power spectrum formalism, this is because
inhomogeneous instrumental noise produces coefficients
of the form $A_{\ell_1 \ell_2}^{LM} \propto {\rm const.}$ 
(in the notation of \cite{Bennett:2010jb}),
which more closely resembles a modulation of the
$\textit{observed}$ power spectrum in real space
than a modulation of the primordial power spectrum
in $k$ space \cite{Bennett:2010jb}. Anomalies are also
seen in estimates of $g_{2M}$ formed from cross correlations
between maps with different noise realizations, which suggests
that noise cannot be the dominant effect \cite{Hanson:2009gu,Bennett:2010jb}.

It has been argued that beams must provide at least a significant
source of bias for estimates of $g(\hat{\mathbf{k}})$ \cite{Hanson:2009gu},
if not a complete explanation \cite{Bennett:2010jb},
although \cite{Groeneboom:2009cb} have also studied the effects of beams
and concluded that they are unimportant.
Here we will address this issue using the new tools of the previous section.
We obtain the coefficients of the instrumental beam
directly from the  
WMAP five-year published beam maps \cite{Hinshaw:2008kr}
by a brute-force
discrete harmonic transform, with beam center determined
simply from the maximum pixel.
For each beam map, we
then scale the resulting $B_{lm}$ such that $B_{00} = 1$,
and average the $A$- and $B$-side beams together, as
appropriate for simulating the effect of differencing
on the final map \cite{Hinshaw:2006ia}.
Finally, we scale these averaged beam transfer functions
at each $\ell$ so that the $B_{l0}$ components agree with
the published WMAP transfer functions.
These are derived from the same TOD observations of
Jupiter which are used to create the beam maps, but do
not suffer from pixelization effects.
Without this scaling, our $B_{l0}$ would still agree with the
published values at a level of better than $1\%$ for
$\ell < 600$.

To get a feeling for the expected effects we begin
by evaluating Eq.~\eqref{eqn:qe_bias_general}.
We only calculate the bias due to the $\Delta^{(1)}$ terms of 
the beam covariance, however we will later verify numerically
that the $\Delta^{(2)}$ terms are not significant for this application, as expected.
As we are ultimately only interested in the low-$\ell$ multipoles
of the beam mean-field it is sufficient to have a model for the
scan strategy $\slm{w}{S}{L}{M}$ at low-$\ell$, and for correspondingly small $S$.
We will initially use $s_{\rm max}\!=\!6$.
To calculate the $\slm{w}{S}{L}{M}$ we use the
analytical method of \cite{Hirata:2004rp}, outlined
in Appendix~\ref{app:toy_scan_strategy}, which
provides an excellent approximation to
the true scan strategy on such large scales.
Following the notation there we use a toy model for WMAP, 
with a spin period of two minutes, a precession period of one hour, and 
scan angles of $\theta_b = 70^{\rm o}$, $\theta_p = 22.5^{\rm o}$.
These are design values for the fiducial center of the WMAP focal plane,
and were achieved with good accuracy in flight \cite{Hinshaw:2003fc}.
The position of the individual detectors within the focal plane does effectively
vary $\theta_b$. We have not corrected for this; however, as we will discuss, 
this does not significantly affect
our results.
The expected mean-field biases are presented in Table~\ref{tab:beam_biases_analytical},
for the $V$- and $W$-band DAs. We initially limit our analysis to $\ell_{\rm max}=400$ for comparison
with earlier results. The inverse-variance filters use
isotropic noise with the appropriate power spectrum. This
evaluation takes only a few seconds for such low $s_{\rm max}$.
The fast precession of the WMAP spin axis gives the scan-strategy azimuthal symmetry in ecliptic coordinates, which
makes $\slm{w}{S}{L}{M} \propto \delta_{M0}$. Because the
bias is proportional to this quantity, it also has this azimuthal
symmetry, which would explain the planar structure of the detected
modulation, and the alignment of the detected effects with the ecliptic poles.
The north-south symmetry of the scan strategy also restricts $\slm{w}{S}{L}{M}$
to even-$L$.
The mean-fields for $\ell=2$ are predicted to be large, and
detectable at many sigma.
Higher multipoles ($\ell=4,6$)
receive much less significant contributions from beams, and
are also not observed to be anomalous in the data \cite{Hanson:2009gu}.
Thus, beams seem to be a likely explanation for the detected anomaly.

Before we move on to the analysis of the WMAP data itself, we
consider some of the insights which the analytic approach makes
possible. Because the detected anomaly is quadrupolar, it depends
on the scan strategy only up to $S=2$ (as there are no $L=2$ modes
for larger $S$). This means that if beams are the explanation they
can only be sourced by the beam's dipole and ellipticity components.
These modes are well constrained by the beam maps, so we expect
our calculations to be quite accurate. As already mentioned, the effective
$\theta_b$ which we use for our scan-strategy calculation differs for some
detectors due to their position in the focal plane. We find that 
$\partial (\slm{w}{2}{2}{0}) / \partial \theta_b = 2\% / {\rm deg.}$ about the fiducial value. 
As all of the $V$- and $W$-band detectors are clustered
within $1^{\rm o}$ in azimuth of the center of the focal plane, we expect
at most $2\%$ errors in the scan-strategy coefficients. This corresponds
to $< 0.2\sigma$ effects for the biases which we have derived.
The analytical calculations also reveal that the dominant contribution
to the quadrupolar bias is given by the $S=2$ modes, rather than
$S=1$. For our purposes, this is quite fortuitous-- there are a number
of opportunities in this calculation for $180\degree$ errors in the
beam orientation angle, however any such errors will not significantly 
affect our results.

Another question which can be asked is whether it might be possible
to verify the effects of beams with an estimator more optimally designed
to detect them.
A Fisher-matrix calculation shows that the $g_{2M}$ estimator
has a typical correlation coefficient of $0.9$
with the optimal quadrupole ``scan-strategy'' estimator with the known WMAP beams,
and so the $g_{2M}$ estimator is effectively optimized to detect beam effects.

Finally, it is interesting to note that the bias for the $Q$-band 
data has the opposite sign of that for the $V$ and $W$ bands.
This effect has already been observed in the data by \cite{Groeneboom:2009cb},
although without explanation.
It is due to the fact that the semimajor axis of the beam ellipticity
for the $Q$ band is oriented parallel to the scan direction, while
the $V/W$ axes are perpendicular \cite{Page:2003eu}. 
A $90^{\rm o}$ rotation corresponds to a sign flip for
the $S=2$ modes of the beam which dominate the bias.
This feature provides strong evidence that the dominant
effect which sources the quadrupolar effect is beam asymmetries.

\begin{table}
\begin{center}
\begin{tabular}{ l c c c }
\hline
\hline
DA
& $\langle g_{20} \rangle$ ($\sigma$)
& $\langle g_{40} \rangle$ ($\sigma$)
& $\langle g_{60} \rangle$ ($\sigma$)  \\
\hline
Q1 & -0.33 (12.1) & 0.030 (0.83) & -0.003 (0.07)\\
Q2 & -0.33 (12.3) & 0.029 (0.81) & -0.003 (0.08)\\
V1 & \ 0.17 (6.51) & 0.031 (0.86) & -0.003 (0.07)\\
V2 & \ 0.17 (6.74) & 0.032 (0.92) & -0.002 (0.06)\\
W1 & \ 0.27 (9.10) & 0.043 (1.07) & -0.002 (0.05)\\
W2 & \ 0.31 (9.79) & 0.042 (0.97) & -0.003 (0.06)\\
W3 & \ 0.33 (9.99) & 0.037 (0.85) & -0.002 (0.05)\\
W4 & \ 0.27 (8.63) & 0.045 (0.95) & -0.003 (0.05)\\
\hline
\hline
\end{tabular}
\end{center}
\caption{Analytic predictions for the beam mean-field bias
to the primordial power asymmetry estimator,
with $l_{\rm max}=400$. The significance ($\sigma$)
is given by the mean-field divided by the estimator noise.}
\label{tab:beam_biases_analytical}
\end{table}

We now turn to the question of whether beam effects are
sufficient to explain completely the quadrupole anomaly,
as seems likely from the significance levels in
Table~\ref{tab:beam_biases_analytical}.
It is straightforward to fold asymmetric beam effects directly into
our analysis~\cite{Hanson:2009gu} of the WMAP five-year data~\cite{Hinshaw:2008kr}; 
we have not yet upgraded to the seven-year data, which are very consistent~\cite{Bennett:2010jb}.
In the quadratic estimator formalism which we use here,
the estimator mean-field and Fisher matrix are determined
on the cut sky by Monte Carlo. By incorporating beam
effects into the CMB simulations using Eq.~\eqref{eqn:obs_eff_har},
the Monte-Carlo mean-field will include the contribution from
beams. Note that this approach will remove the subdominant
contribution due to the $\Delta^{(2)}$ terms, as well as the $\Delta^{(1)}$
terms which we used for the analytical calculation, although we
have verified numerically that the $\Delta^{(2)}$ terms do not contribute
more than a few percent to the beam mean-field.
For the convolution we use $s_{\rm max}\!=\!2$, as higher terms
do not contribute to the quadrupolar anisotropy.
To obtain a minimum-variance estimator, we should
also incorporate beam effects into the inverse-variance filtering
operation, as discussed in the previous section,
however we find that the estimator noise variance for the asymmetrically convolved simulations is less than $10\%$ higher than that without
beam effects, and so this improvement would not have a significant
effect on our results. 

\begin{figure}
\begin{center}
\includegraphics[width=\columnwidth]{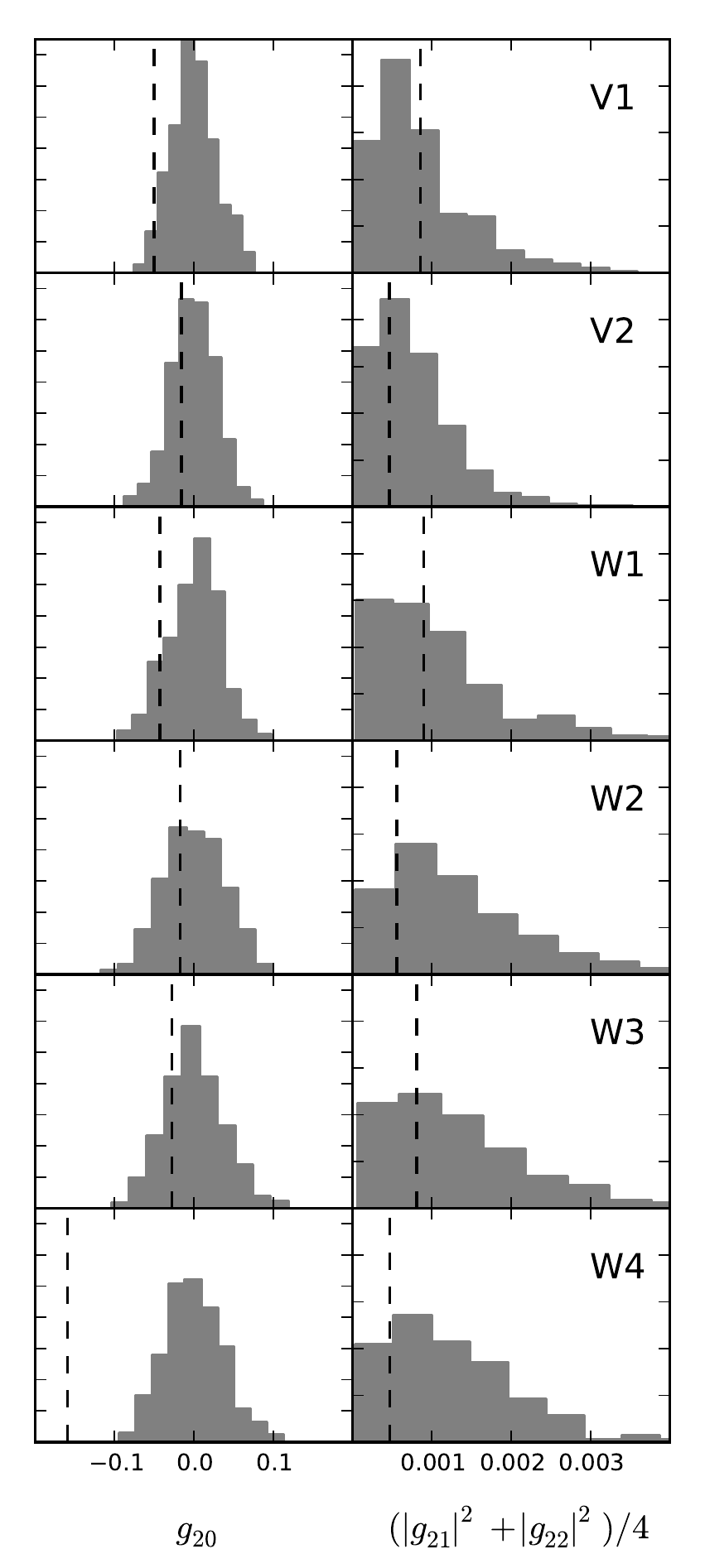}
\end{center}
\caption{
Significance of the observed WMAP primordial-power-modulation quadrupole, with correction for beams,
for the WMAP $V$-band foreground-reduced data, limited to $\ell_{\rm max}=1000$.
This is essentially a beam-corrected version of Fig.~9 in \cite{Hanson:2009gu}.
The gray histograms are from the 500 simulations which are
used to determine the estimator Fisher matrix and mean-field.
The dashed vertical lines are for the observed data.
Detailed interpretation is provided in the text.
}
\label{fig:primordial_handicraft_mf_norm_quad_cl2s}
\end{figure}

The significance of the measured $g_{2M}$ is
plotted in Fig.~\ref{fig:primordial_handicraft_mf_norm_quad_cl2s}
for the foreground-reduced WMAP $V$- and $W$-band DAs,
incorporating all of the usable signal in the WMAP data by taking 
$\ell_{\rm max} = 1000$. 
It can be seen that
the mean-field subtraction of beam effects results in data which are
consistent with the isotropic model. The non-$\delta_{M0}$ values are very
similar for all detectors, as one would expect for an isotropic sky, given that a large fraction of
the estimator ``noise'' is due to the CMB fluctuations themselves, which
are common between detectors. The variation for the $\delta_{M0}$ modes
is more significant, indicating that there may be some small errors in the
mean-field subtraction. The measured value of $g_{20}$ is not significant
in any detector, however, with the exception of $W4$ which shows
a large negative bias even after subtraction of the beam mean-field.
It should be kept in mind that without beam subtraction, each DA shows $6-9\sigma$
effects in the $\delta_{M0}$ modes.
We believe that the residual anomaly in W4 may be attributed to the effects of correlated noise.
The W4 DA has clearly correlated noise even after the prewhitening stage
of the WMAP analysis \cite{Hinshaw:2003fc}. Its $1/f$ knee frequency
is several times larger than any of the other DAs \cite{Jarosik:2003fe}.
A negative bias to $g_{20}$ from correlated noise is also expected
analytically. For subdominant correlated noise, the effects of striping
can be modeled as a convolution of the pixel-uncorrelated instrumental
noise with a narrow beam which has its semimajor axis along the scan
direction. As we have already seen with the $Q$-band data, this leads
to a negative bias in $g_{20}$. 
We can further test this hypothesis by forming our quadratic estimator
from pairs of maps with uncorrelated noise. For this we use the $W4$ data
for individual years. Autocorrelating the data for any single year and correcting
for beam effects, we continue to find a large negative bias in $g_{20}$
(albeit with slightly less statistical significance due to the larger instrumental
noise for only a single year of data). Cross-correlating data from any pair of
separate years, however, the effect disappears.
Although we do not plot them here, the $Q$-band data
are also completely consistent with $g_{20}=0$ after correction for
beam effects.
We therefore assert that beam effects provide a sufficient explanation for the detected anomaly.
Based on the average diagonal elements of our simulation Fisher matrices,
we place a conservative limit on any single mode of $|g_{2M}| <  0.07$ 
at $95\%$ confidence. The corresponding
limit for the power spectrum of $g_{2M}$ is  $C^{gg}_2 < 0.003$.

\textit{Disagreement with other results.--}
In \cite{Hanson:2009gu} we tested the effect of beam 
asymmetries using the simulations of 
\cite{Wehus:2009zh}, 
which appeared only partially to explain the strong 
detected signal. However, on closer inspection, we 
find problems with these simulations, indicating that 
they are not representative of the effect of the WMAP beams. 
Applying the modulation estimator to these 
simulations with no noise, we find that they show a strong mean-field,
with a spatial pattern which is identical between detectors, as expected
analytically. The mean-field does not, however, have the purely planar structure
associated with a $\delta_{M0}$ pattern in ecliptic coordinates.
Such an error is most likely due to 
errors in the beam orientation angles $\alpha_i$;
however, private communication with Eriksen et. al.\
has been unsuccessful in revealing the precise origin 
of this discrepancy.
Our results are also discrepant with \cite{Groeneboom:2009cb},
who analyzed one of the 
simulations 
from \cite{Wehus:2009zh}
and found no beam effects.
We believe this is due to two factors:
(i) the simulation which they analyzed
is one in which we also saw small effects
in our original work \cite{Hanson:2009gu}; and
(ii) the nonplanar structure of the mean-field in the 
simulations poses
difficulty for the estimator used by \cite{Groeneboom:2009cb}, which searches
explicitly for the planar mode.

\section{Conclusions}
\label{sec:conclusions}
Beam asymmetry effects can be more important
for anisotropy estimators than for power spectrum
analysis, because their effects generally enter at lower order
in the asymmetry.  
Beam asymmetries fit nicely into 
the larger formalism of quadratic anisotropy estimators.
They result in a mean-field bias which directly traces the scan strategy 
$\slm{w}{S}{L}{M}$, and can be calculated analytically 
on the full sky or determined from Monte-Carlo simulations on the
cut sky.

Beam effects appear to provide a sufficient explanation
for the $9\sigma$ detection of an apparent quadrupolar
modulation of the primordial power spectrum in the WMAP data.
We note that the WMAP team already incorporates
the effects of beam asymmetry into their power
spectrum analysis (where it is a much smaller effect
in any case), and so the resolution of this anomaly
should not have any effect on their cosmological
parameter constraints.

All of this work will apply directly to the Planck
experiment, which has a less symmetrizing scan strategy
than WMAP. Planck's
increased sensitivity also opens up the field for the precision
analysis of interesting astrophysical secondaries such as
the anisotropic signal from gravitational lensing. 
The tools and techniques which we have discussed here
may also be extended straightforwardly for use with polarization data.

\section{Acknowledgements}
DH is grateful for the support of a Gates Scholarship; AL acknowledges a PPARC/STFC Advanced Fellowship.
Some of the results in this paper have been derived using
HEALPix~\cite{Gorski:2004by}.
We acknowledge the use of the Legacy Archive for Microwave Background
Data Analysis (LAMBDA).
Support for LAMBDA is provided by the NASA Office of Space Science.
We thank Kendrick Smith, David Spergel, and Jo Dunkley for helpful comments on a draft version of this manuscript, as
well as Hans Kristian Eriksen and Nicolaas Groeneboom for correspondence.
\appendix


\section{Toy Scan Strategy}
\label{app:toy_scan_strategy}
Here we consider a simple model of a scan strategy
which serves as a good approximation to typical satellite experiments.
It consists of:
\begin{enumerate}[(1)]
\item a beam at an angle $\theta_b$ to the satellite spin axis, which rotates with period $\tau_s$;
\item a precession at an angle $\theta_p$ to the antisolar direction, with period $\tau_p$; and
\item a continuous repointing of the antisolar direction as the observer orbits the sun.
\end{enumerate}
If $\tau_s \ll \tau_p \ll 1\,{\rm year,}$ then $w(\Omega_p,s)$ can be calculated analytically \cite{Hirata:2004rp}.
First, we calculate the quantity
$v(\Omega_p,s) = \sum_{i \in p} e^{i s \alpha_i}$
and then we form $w(\Omega_p, s) = v(\Omega_p ,s)/v( \Omega_p,0)$.

To calculate $v(\Omega_p,s)$, begin in a coordinate system which
places the spacecraft spin axis along the $+z$ axis.
Rotation about the spin axis in these coordinates gives
$v^{(1)}(\Omega_p, s) \propto \delta(\theta - \theta_b) e^{is0}$. Expanding
this using the appropriate spin harmonics gives
\be
[v^{(1)}(\Omega_p, s)]_{lm} = K \delta_{m0} \,\yslm{s}{l}{0}(\theta_b, 0),
\ee
where $K$ is some constant.
We can then rotate out to place the precession axis along
the $+z$ axis, obtaining $v(p,s)$ in precession coordinates:
\be
[v^{(2)}(\Omega_p, s)]_{lm} = \sum_{m''} D^{l}_{m m''}(0, \theta_p, 0) [v^{(1)}(\Omega_p, s)]_{lm''} .
\ee
Rotation about the precession axis removes all but the $m=0$ components giving
\ba
[v^{(2)}(\Omega_p, s)]_{lm}
&=& \delta_{m 0} D^{l}_{0 0}(0, \theta_p, 0) [v^{(1)}(\Omega_p,s)]_{l0} \nn \\
&=& \delta_{m 0} K P_l(\cos\theta_p) \yslm{s}{l}{0}(\theta_b, 0).
\ea
Finally, the precession axis is rotated $90\degree$ down to the
ecliptic plane, and again only $m=0$ modes in the new coordinates
are taken, to effect the azimuthal averaging given by the yearly
rotation about the sun:
\be
[v(\Omega_p, s)]_{lm} = \delta_{m 0} K P_l(0) P_l(\cos\theta_p) \yslm{s}{l}{0}(\theta_b, 0).
\label{eqn:vps_analytical}
\ee
This can be used to calculate $w(\Omega_p,s)$ in ecliptic coordinates.
Note that only multipoles with $l$ even are nonzero due to the north-south
symmetry of the scan pattern. This is enforced by the dependence on $P_l(0)$.




\section{Beam asymmetries and power spectra}
\label{sec:beam_power}

For the CMB temperature power spectrum, beam asymmetries are only important
at high $l$ (i.e.\ below the beam scale). On such scales, a pseudo-$C_l$
analysis is usually employed in which a weight function is applied to
the observed sky and the empirical (pseudo) power spectrum of the weighted sky
is taken. The expectation value of the pseudo-$C_l$, after removal of the
bias due to instrument noise, is linearly related to the true power
spectrum, $C_l^{\Theta\Theta}$. In this Appendix, we calculate this relation
in the presence of beam asymmetries; for related work
see~\cite{Hinshaw:2006ia}. Good performance can be obtained from
pseudo-$C_l$ estimators with a careful
choice of weight function~\cite{Efstathiou:2003dj}, such as a local
approximation to the optimal inverse signal-plus-noise weighting. Below the
beam scale, the signal is exponentially suppressed and weighting by the inverse
variance of the pixel noise is close to optimal. 

We noted in Sec.~\ref{sec:covariance} that, for full-sky observations
and uniform weighting of the data, beam asymmetries only affect the power
spectrum at second order. This is no longer true with anisotropic weighting,
which generally will arise from the survey geometry or inhomogeneities in the
noise. In this case, we generalize Eq.~(\ref{eq:spin_weight}) to include
a spin-0 pixel weight function on the right-hand side.
Equation~(\ref{eqn:obs_eff_har}) then still holds, but ${}_0 w_{LM}$ is
no longer necessarily zero for $L\neq 0$.
Writing the integral of three spin harmonics
in terms of $3j$ symbols, we have
\begin{align}
\tilde{\Theta}_{l'm'} &= \sum_{LMS}\sum_{lm}(-1)^{m'} \sqrt{\frac{(2l+1)(2l'+1)
(2L+1)}{4\pi}} \nonumber \\
&\hspace{-1cm} \times \threej{l'}{l}{L}{-m'}{m}{M} \threej{l'}{l}{L}{0}{-S}{S}
B_{lS} {}_{-S}w_{LM}\Theta_{lm} .
\end{align}
Inserting this in the definition of the pseudo power spectrum,
\begin{equation}
\tilde{C}_l \equiv \frac{1}{2l+1} \sum_{m} |\tilde{\Theta}_{lm}|^2 ,
\end{equation}
and taking the expectation value, we find
\begin{align}
\langle \tilde{C}_{l'} \rangle &= \sum_{lS'LS}\frac{(2l+1)(2L+1)}{4\pi}
\threej{l'}{l}{L}{0}{-S}{S} \threej{l'}{l}{L}{0}{-S'}{S'} \nonumber \\
&\mbox{}\hspace{1cm}
\times B_{lS}B^*_{lS'} {}_{(-S\,-S')}\mathcal{W}_L C_l^{\Theta\Theta},
\label{eq:pCl}
\end{align}
where we have used an orthogonality relation for the $3j$ symbols. Here,
the scan strategy and weight function are encoded in the cross spectra
\begin{equation}
{}_{(S\,S')}\mathcal{W}_L \equiv \frac{1}{2L+1}\sum_{M} {}_{S}w_{LM}
{}_{S'}w^*_{LM} .
\end{equation}
For mildly asymmetric beams, or a wide distribution of crossing angles,
the sums over $S$ and $S'$ in Eq.~(\ref{eq:pCl}) can be truncated after
only a few terms. In this case, the effect of beam asymmetries on the
mean pseudo-$C_l$ can be computed efficiently with no further simplifying
assumptions.

For the case of symmetric beams ($B_{lS} = \delta_{S0} B_{l0}$), or for
a uniform distribution of observation angles in each pixel
(${}_S w_{LM} = \slm{w}{S}{L}{M} \delta_{S0}$),
Eq.~(\ref{eq:pCl}) reduces to the usual result~\cite{Efstathiou:2003dj}
for symmetric beams.

\vfill

\bibliography{beams}

\appendix

\end{document}